\documentclass[12pt]{article}
\usepackage{amsmath}
\usepackage{graphicx,color}
\usepackage{setspace}
\begin{document}
%\large

%\begin{frontmatter}
\title{Verification of generalized Kramers-Kr\"{o}nig relations and sum rules on experimental data of third harmonic generation susceptibility on polymers}
%[Verification of K-K relations and sum rules for $\chi^{(3)}(3\omega)$ on polymers]
\author{\textbf{Valerio Lucarini} \\ \and \textbf{Kai-Erik Peiponen} \\ Department of Physics University of Joensuu \\ FIN-80101 Joensuu Finland}
%\affiliation{Department of Physics \\ University of Joensuu \\
%FIN-80101 Joensuu Finland}
%\author{Kai-Erik Peiponen}
%\address{Department of Physics \\ University of Joensuu \\ FIN-80101 Joensuu Finland}
\maketitle
%\centering{Department of Physics, University of
%Joensuu, FIN-80101 Joensuu Finland}
\begin{abstract}
We present the first analysis of harmonic generation data where
the full potential of the generalized nonlinear Kramers-Kr\"{o}nig
(K-K) relations and sum rules is exploited. We consider two
published sets of wide spectral range experimental data of third
harmonic generation susceptibility on different polymers, the
polysilane (frequency range: $0.4-2.4$ $eV$) and the polythiophene
(frequency range: $0.5-2.0$ $eV$). We show that, without extending
the data outside their range with the assumption of an a-priori
asymptotic behavior, independent truncated dispersion relations
connect the real and imaginary part of the moments of the third
harmonic generation susceptibility
$\omega^{2\alpha}\chi^{(3)}(3\omega; \omega, \omega ,\omega)$,
with $\alpha$ ranging from 0 to 3, in agreement with the theory,
while there is no convergence if we choose $\alpha=4$. We repeat
the same analysis for $\omega^{2\alpha}[\chi^{(3)}(3\omega;
\omega, \omega ,\omega)]^{2}$ and show that a larger number of
independent K-K relations connect the real and the imaginary part
of the function under examination. We also compute the sum rules
for the suitable moments of the real and imaginary parts, and
observe that only considering higher powers of the susceptibility
the correct vanishing sum rules are more precisely obeyed. All our
results are in fundamental agreement to recent theoretical
findings. Sum rules providing explicit information about
structural properties of the material seem to require wider
spectral range. These constraints are expected to hold for any
material and provide fundamental tests of self-consistency that
any experimental or model generated data have to obey; similar
tests of coherence can be performed for other nonlinear optical
processes, e.g. pump-and-probe. Verification of K-K relations and
sum rules constitute unavoidable benchmarks for any investigation
that addresses the nonlinear response of matter to radiation on a
wide spectral range.

\end{abstract}
%\end{frontmatter}

\clearpage

\section{Introduction}\label{sect1}

Kramers-Kr\"{o}nig (K-K) dispersion relations and sum
rules\cite{Landau,Kaibook,Altarelli1,Altarelli2} have constituted
for a long time fundamental tools in the investigation of
light-matter interaction phenomena in condensed matter, gases,
molecules, and liquids because they provide constraints able to
check the self-consistency of experimental or model-generated data
\cite{Shiles et al. 1980}. In particular these general properties
allow to frame peculiar phenomena related to matter-matter or
light-matter coupling  that are very relevant at given
frequencies, such as excitonic or polaritonic effects in
solids\cite{Grosso}, in the context of the interaction on the
whole spectral range, showing that their dispersive and absorptive
contributions are connected to all the other contributions in the
rest of the spectrum\cite{Altarelli3}. Moreover by applying K-K
relations it is possible to invert optical data, i.e. acquiring
knowledge on dispersive phenomena by measurements over all the
spectrum of absorptive phenomena or viceversa
\cite{Altarelli3,Aspnes}. The conceptual foundations of such
general properties, is the principle of causality\cite{Milonni} in
the light-matter interaction, and the Titmarsch's
theorem\cite{Nussenzveig} provides the connection between the
mathematical properties of the functions describing the physics in
the time and frequency domains.
\par
In spite of the ever increasing scientific and technological
relevance of optical harmonic generation processes, and in general
of nonlinear optical phenomena, relatively little attention has
been paid to experimental investigation of the general properties
of the corresponding nonlinear susceptibilities\cite{Lee et al.
85,Cataliotti 97,Gubler et al. 2000,Martinelli et al. 2000},
because the research has usually focused on achieving high
resolution in both experimental data and theoretical calculations.
Rapidly developing technologies of tunable lasers allow to span
larger and larger spectral ranges, so that the experimental
investigations of frequency dependent nonlinear optical properties
of matter is becoming more and more possible. In the context of
this new class of experiments, K-K relations and sum rules could
provide information on whether or not a coherent, common picture
of the nonlinear properties of the material under investigation is
available.
\par
The first heuristic applications of K-K theory to nonlinear
susceptibilities date back to the '60 \cite{Price,Kogan,Caspers},
while a more systematic study has begun essentially in the last
decade. Some authors have preferentially introduced the K-K  in
the context of ab-initio or model calculations of materials
properties \cite{Ghahramani et al. 1991,Sheik et al. 92,Kador 95,Hughes et al. 1997}.\\
Other authors have used a more general approach able to give
theoretical foundations of dispersion theory for nonlinear optics
\cite{Kai1,Kai2,Bassani1,Scandolo,Bassani2}. The instruments of
complex analysis have allowed the definition of necessary and
sufficient conditions for the applicability of K-K relations,
which require the holomophicity of the susceptibility in the upper
complex plane of the relevant frequency variable, given the form
of the nonlinear susceptibility descriptive of the nonlinear
phenomena under examination. It has also been shown that the
asymptotic behavior of the nonlinear susceptibility, which can be
obtained using the nonlinear Kubo-response function
formalism\cite{Kubo}, determines the number of independent pairs
of K-K relations that hold simultaneously. In the cases K-K do not
hold, other techniques can be applied for the inversion of optical
data \cite{Kaibook,Peiponen02}. Combining K-K relations and
knowledge of the asymptotic behavior of the nonlinear
susceptibility sum rules for nonlinear optics, obtained by other
authors by following different strategies\cite{Chernyak}, can be
naturally derived along the same lines of the linear case.
\par
Recent theoretical advances\cite{ValerioHm,ValerioH} have
generalized the results obtained for the second \cite{Bassani2}
and third harmonics \cite{Rapapa} by showing arbitrary order
harmonic generation susceptibilities $\chi^{(n)}(n\omega; \omega,
\cdots ,\omega)$ are holomorphic in the upper complex $\omega$
plane, and asymptotically decrease as $\omega^{-2n-2}$; therefore
K-K relations hold for all the moments
$\omega^{2\alpha}\chi^{(n)}(n\omega; \omega, \ldots ,\omega)$
(from now on instead of $\chi^{(n)}(n\omega; \omega, \ldots
,\omega)$ we use the simpler notation $\chi^{(n)}(n\omega)$) with
$0\leq\alpha\leq n$. From these a total of \textit{2n+2} sum rules
for the moments of the real and imaginary part of the
susceptibilities can be derived, only one (the $(2n+1)^{th}$ of
the imaginary part) being different from zero and descriptive of
the inner structure of the material under investigation. The
previous results have been recently extended \cite{jarkko} to the
wider class of the moments of the $k^{th}$ powers of harmonic
generation susceptibility, which can be written as
$\omega^{2\alpha}[\chi^{(n)}(n\omega)]^{k}$. In this case K-K
relations hold if $0\leq\alpha\leq k(n+1)-1$, and therefore a
total of $2k(n+1)-2$ sum rules can be derived, with the
$(2k(n+1)-1)^{th}$ moment of imaginary part giving the only
nonzero summation over all the spectrum. The fundamental reason of
this results is that if the harmonic generation susceptibility is
holomophic in the upper complex $\omega$ plane, so are its powers.
Higher powers of the susceptibility have faster asymptotic
decrease, so that the limitations related to the finite range of
real data are expected to be relaxed: this should be of particular
benefit to the convergence of the sum rules, for which the
consideration of the asymptotic behavior is more critical than
for K-K relations \cite{Fano,Altarelli3,Bachelet}. \\
The consideration of anchor points\cite{Ahrenkiel,Palmer} in the
unknown part of $\chi^{(n)}(n\omega)$\cite{ValJarKai} is a very
promising technique for improving data inversion. \\ Until now,
only in few investigations independent measurements of the real
and imaginary parts of the harmonic generation susceptibilities
have been performed for a relatively wide
range\cite{Torruellas1,Torruellas2,Kishida,Kishida2}, and
consequently  the verification of the coherence of measured data
by check of the self-consistency of K-K relations is still of very
limited use\cite{Kishida}.
\par
In this paper we present the first analysis of harmonic generation
data where the full potential of the generalized nonlinear K-K
relations and sum rules for harmonic generation susceptibilities
is exploited. We consider two published sets of wide spectral
range experimental data of third harmonic generation
susceptibility on different polymers, the polysilane
\cite{Kishida} (frequency range: \textit{0.4 - 2.5 eV}) and the
polythiophene \cite{Torruellas1} (frequency range: \textit{0.5 -
2.0 eV}). \\ In section $2$ we apply K-K transformations to the
real and imaginary part of the moments of the third harmonic
generation susceptibility $\omega^{2\alpha}\chi^{3}(3\omega)$,
with a suitable choices of $\alpha$, and describe the quality of
the data inversion obtained in this way. We also perform the same
analysis for $\omega^{2\alpha}[\chi^{3}(3\omega)]^{2}$, in order
to prove for the first time on experimental data that K-K
relations hold also for the powers of the susceptibility, and show
how many additional independent dispersion relations connect the
real and the imaginary part.
\\ In section $3$ we compute the sum rules for the experimental
data, by calculating the integrals of the suitable moments of the
real and imaginary parts of the susceptibility and of the $k^{th}$
power of the susceptibilities, with $1\leq k\leq5$, and present a
discussion on the issue of their convergence. \\
In section $4$ we present our conclusions.

\section{Efficacy of generalized K-K relations for data inversion}\label{sect2}

In this study we base our calculations on two published sets of
experimental data on third harmonic generation on polymers, where
the real and imaginary part of the susceptibility were
independently measured. The first data set refers to measurements
taken on polisylane \cite{Kishida} and spans a frequency range of
\textit{0.4 - 2.5 eV}. The second data set refers to measurements
taken on polythiophene \cite{Torruellas1} and spans a frequency
range of \textit{0.5 - 2.0 eV}. \\ In this paper we control the
self-consistency of the two data sets by observing the efficacy of
the K-K relations in inverting the optical data  for the functions
$\omega^{2\alpha}[\chi^{3}(3\omega; \omega, \omega ,\omega)]^{k}$,
with k=1,2. The theory\cite{ValerioHm,ValerioH,Rapapa,jarkko}
prescribes convergence of the dispersion relations:
\begin{equation}\label{k3}
\Re\{[\chi^{(n)}(n\omega')]^{k}\}=\frac{2}{\pi\omega'^{2\alpha}}\wp\int_{0}^{\infty}\frac{\omega^{2\alpha+1}\Im\{[\chi^{(n)}(n\omega)]^{k}\}}{\omega^{2}-\omega'^{2}}{\rm{d}}\omega,
\end{equation}
\begin{equation}\label{k4}
\Im\{[\chi^{(n)}(n\omega')]^{k}\}=-\frac{2}{\pi\omega'^{2\alpha-1}}\wp\int_{0}^{\infty}\frac{\omega^{2\alpha}\Re\{[\chi^{(n)}(n\omega)]^{k}\}}{\omega^{2}-\omega'^{2}}{\rm{d}}\omega,
\end{equation}

only if $0\leq \alpha \leq k(n+1)-1$, therefore in our case of
$n=3$ K-K relations should work only if $0\leq \alpha \leq 3$ if
$k=1$ and $0\leq \alpha \leq 7$ if we consider the second power of
the susceptibility. In our study we do not assume any asymptotic
behavior outside the data range, but use only the experimental
data, because extrapolation is somewhat arbitrary and in K-K
analysis can be quite problematic \cite{Kai3,Aspnes}; we
effectively apply truncated K-K relations and use a
self-consistent procedure.
\par
In the paper by Kishida \textit{et al.} \cite{Kishida} a check of
validity of the K-K relations  was already performed by comparing
measured and retrieved $\chi^{(3)}(3\omega)$. Apart from the fact
that our analysis considers also another data set, the
consideration of the moments of the susceptibility is not a mere
add-on to the work by Kishida \textit{et al.}\cite{Kishida}, but
it represents a fundamental conceptual improvement: these
additional independent relations are peculiar to the nonlinear
phenomena, and provide independent double-checks of the
experimental data that must be obeyed in addition to the
conventional K-K relations.
\par
\begin{figure}
  \includegraphics[angle=90,width=\textwidth]{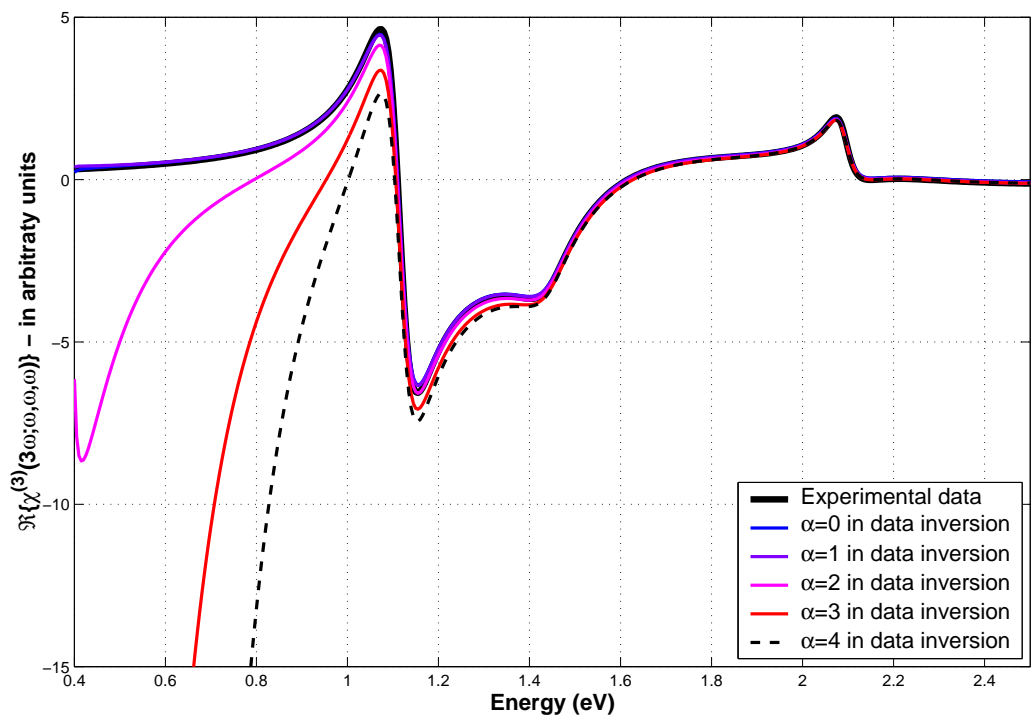}\\
  \caption{Lucarini, The Journal of Chemical Physics}\label{f1}
\end{figure}
\begin{figure}
  \includegraphics[angle=90,width=\textwidth]{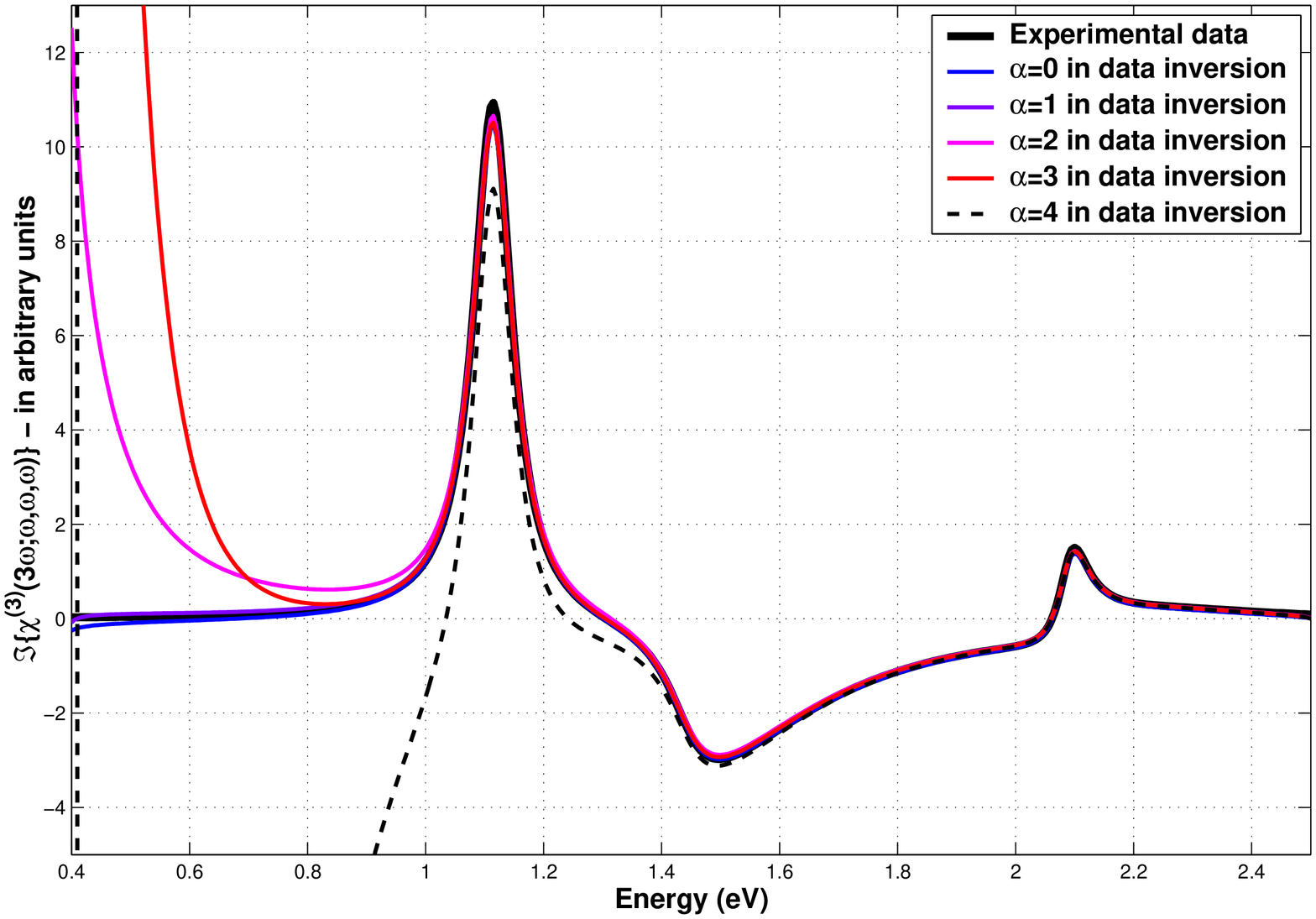}\\
  \caption{Lucarini, The Journal of Chemical Physics}\label{f2}
\end{figure}

We present in figures \ref{f1} and \ref{f2} the results of K-K
inversion for respectively the real and imaginary part of the
third harmonic generation susceptibility data on polysilane. We
observe that in both cases the retrieved data obtained with the
choices $\alpha=0,1$ are almost indistinguishable from the
experimental data, while for $\alpha=2$ and $\alpha=3$, the
agreement is quite poor in the lower part of the spectrum: the
error induced by the presence of the cut-off at the high frequency
range becomes more critical in the data inversion for larger
$\alpha$, since a slower asymptotic decrease is realized. We
expect that inverting the data with the additional information
given by anchor points located in the lower part of the data range
these divergences can be cured. However, here our object is to
deal with the worst case i.e. there is no a priori information
about the phase of the complex nonlinear susceptibility at one or
more fixed anchor points (fixed angular frequencies)
\cite{ValJarKai}. From the theory we expect that for $\alpha=4$ no
convergence should occur: actually we observe that, while the main
features around $1.1$ $eV$ are represented, there is no
convergence at all for lower frequencies; the absence of a clear
transition in retrieving performance between the $\alpha=3$ and
the $\alpha=4$ case is due to the finiteness of the data range.
\par
\begin{figure}
  \includegraphics[angle=90,width=\textwidth]{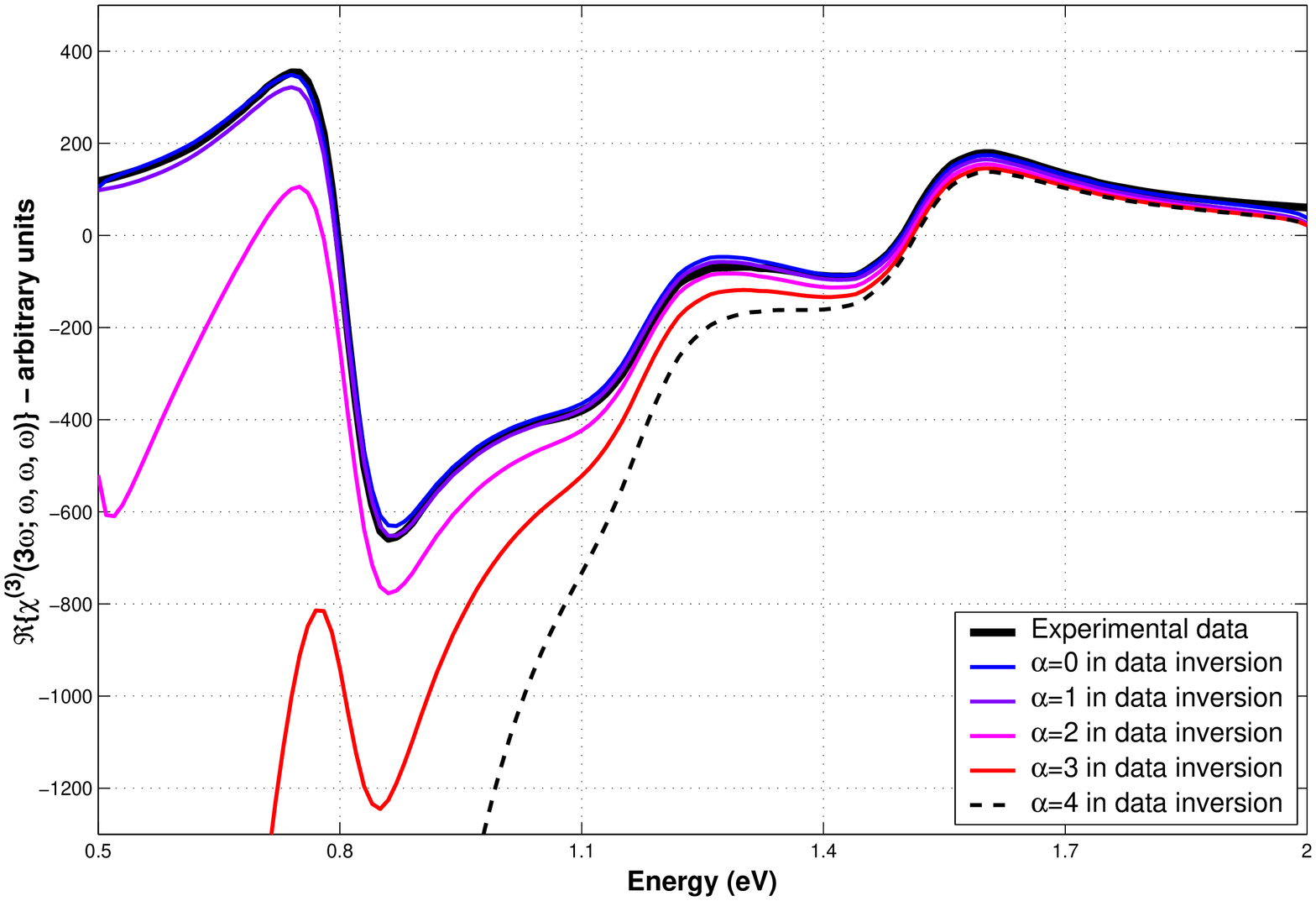}\\
  \caption{Lucarini, The Journal of Chemical Physics}\label{f3}
\end{figure}
\begin{figure}
  \includegraphics[angle=90,width=\textwidth]{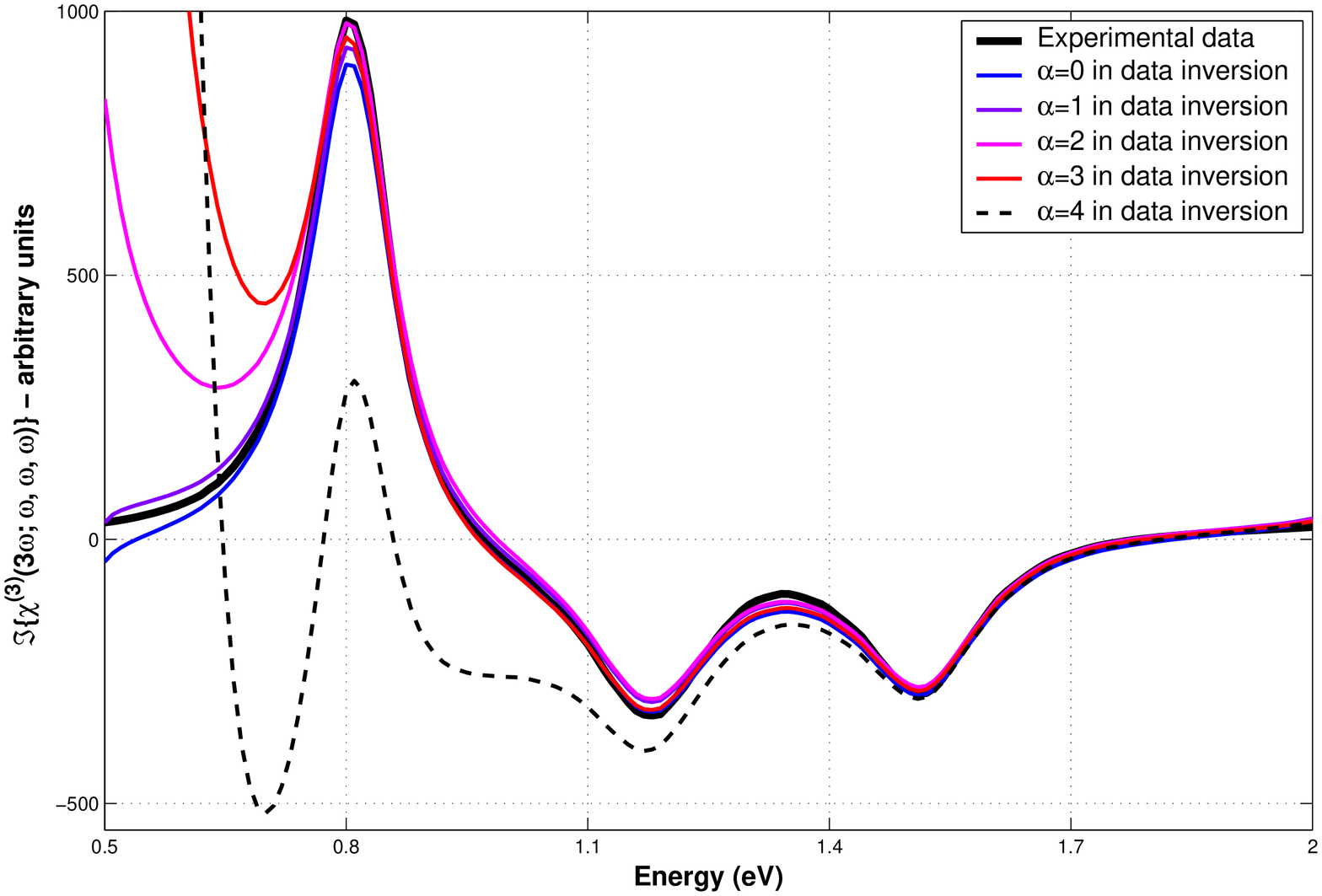}\\
  \caption{Lucarini, The Journal of Chemical Physics}\label{f4}
\end{figure}

In figures \ref{f3} and \ref{f4} we show the comparison between
retrieved and experimental data of third harmonic generation
susceptibility on polythiophene for the real and imaginary part
respectively. The dependence of the accuracy of quality of data
inversion is similar to the previous case: for $\alpha=0,1$ the
agreement is virtually perfect, while for $\alpha=2,3$ we have
progressively worse performance in the low frequency range; anyway
the peaks in the imaginary part are still well reproduced, while
the dispersive structures in the real part are present but shifted
towards lower values. In this case the quality of retrieved data
for $\alpha=4$ is more distinct from what obtained with $\alpha=3$
than in the previous data set. The inversion with $\alpha=4$
presents a notable disagreement in the whole lower half of the
data range for both real and imaginary part, in particular we see
that in figure \ref{f3} the dispersive structure is absent, while
the main peak in figure \ref{f4} is essentially missed. \par
Usually it is likely to expect that only the real or the imaginary
part of the nonlinear susceptibility has been measured. Then
normal procedure is to try data inversion using K-K in order to
calculate the missing part.
\par
The results on figures $1-4$ confirm that best convergence is
obtained when using conventional K-K, therefore these should
generally be used to obtain a \textit{first best guess} for the
inversion of optical data, and should be used as seed for any
self-consistent retrieval procedure; nevertheless if there is good
agreement with the inversions obtained with higher values of
$\alpha$, it is reasonable to conclude that the dispersion
relations provide much more robust results. In this sense, the two
data sets here analyzed are quite good in terms of
self-consistency.
\par
We underline that if on one side considering a higher power of the
susceptibility implicitly filters out noise and errors in the
tails of the data, on the other side experimental errors in the
relevant features of the spectrum, peaks for the imaginary part
and dispersive structures for the real part, are greatly enhanced
if higher powers of the susceptibility are considered; in the
latter case consistency between K-K inversion of different moments
is expected to be more problematic than in the $k=1$ case.
Therefore improved convergence for more moments will occur for the
powers of the susceptibility $k>1$ \textit{only if} the data are
basically good.
\par
\begin{figure}
  \includegraphics[angle=90,width=\textwidth]{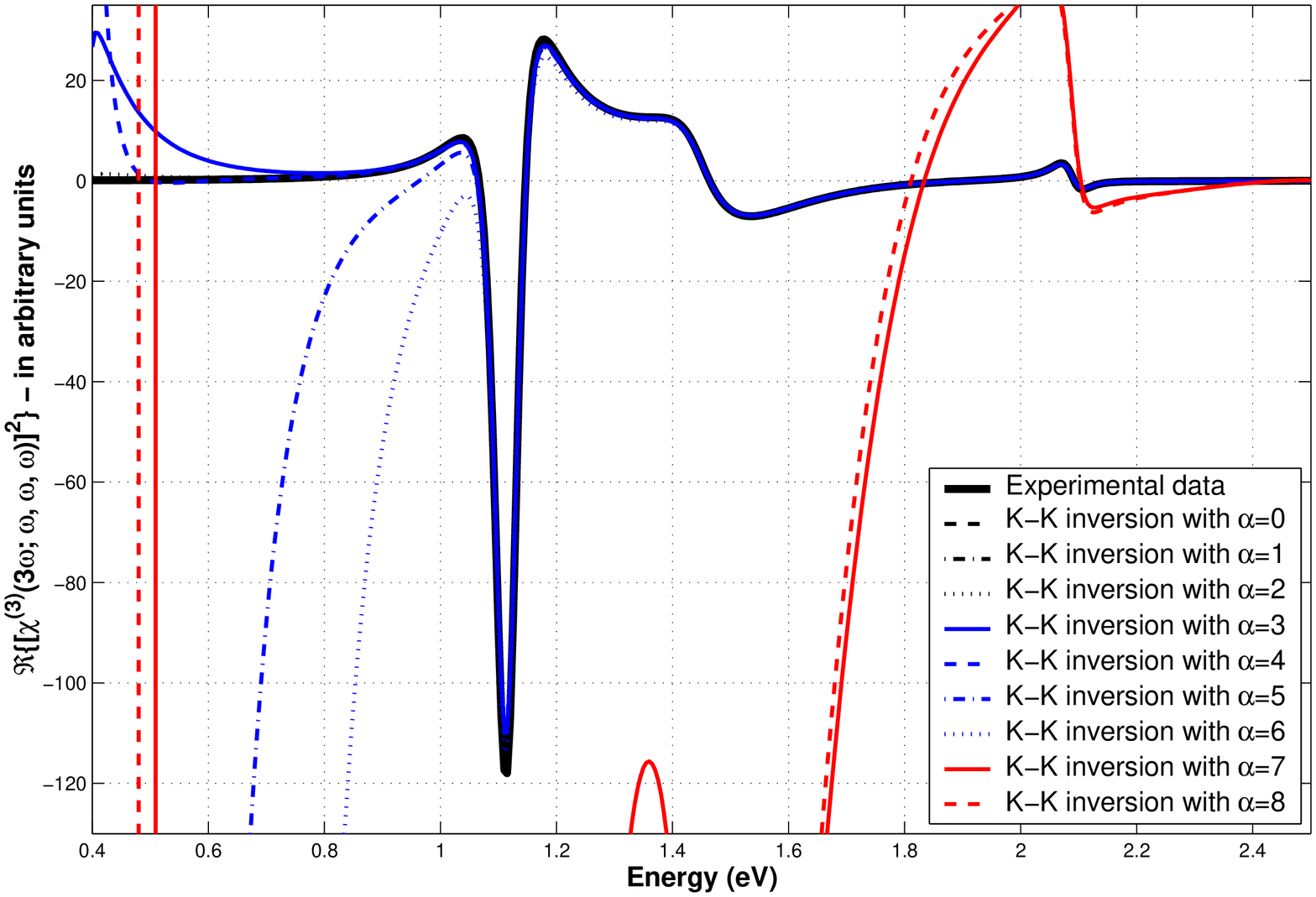}\\
  \caption{Lucarini, The Journal of Chemical Physics}\label{f5}
\end{figure}
\begin{figure}
  \includegraphics[angle=90,width=\textwidth]{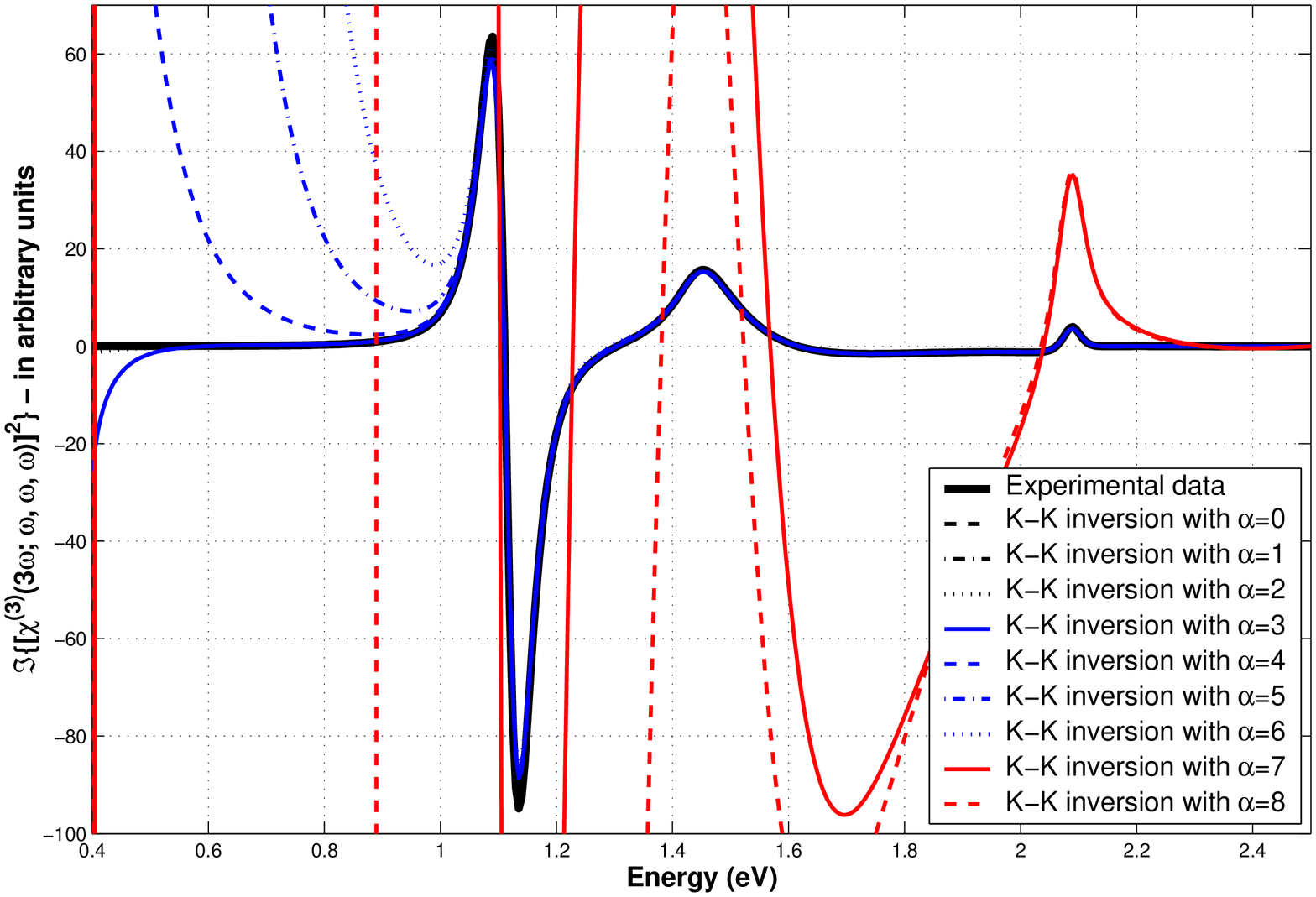}\\
  \caption{Lucarini, The Journal of Chemical Physics}\label{f6}
\end{figure}

In figures \ref{f5} and \ref{f6} we show the
results of K-K inversion for respectively the real and imaginary
part of the second power of the third harmonic generation
susceptibility on polysilane. Up to our knowledge this is the
first analysis of this kind on experimental data. We observe that
for $\alpha=0,1,2$ the agreement between experimental and
retrieved data is almost perfect, while it gets progressively
worse for increasing $\alpha$. Nevertheless as long as $\alpha
\leq 6$ the main features are well reproduced for both the real
and imaginary part and the retrieved data match well if the photon
energy is $\geq$ $1.0$ $eV$. The theory predicts convergence for
$\alpha=7$ and divergence for $\alpha=8$: in our analysis we have
divergence also in the former case, and it is reasonable to
attribute this to the cut-off in the high frequency range of the
data, because a very high moment as the $14^{th}$ requires a very
well defined asymptotic behavior.
\par
\begin{figure}
  \includegraphics[angle=90, width=\textwidth]{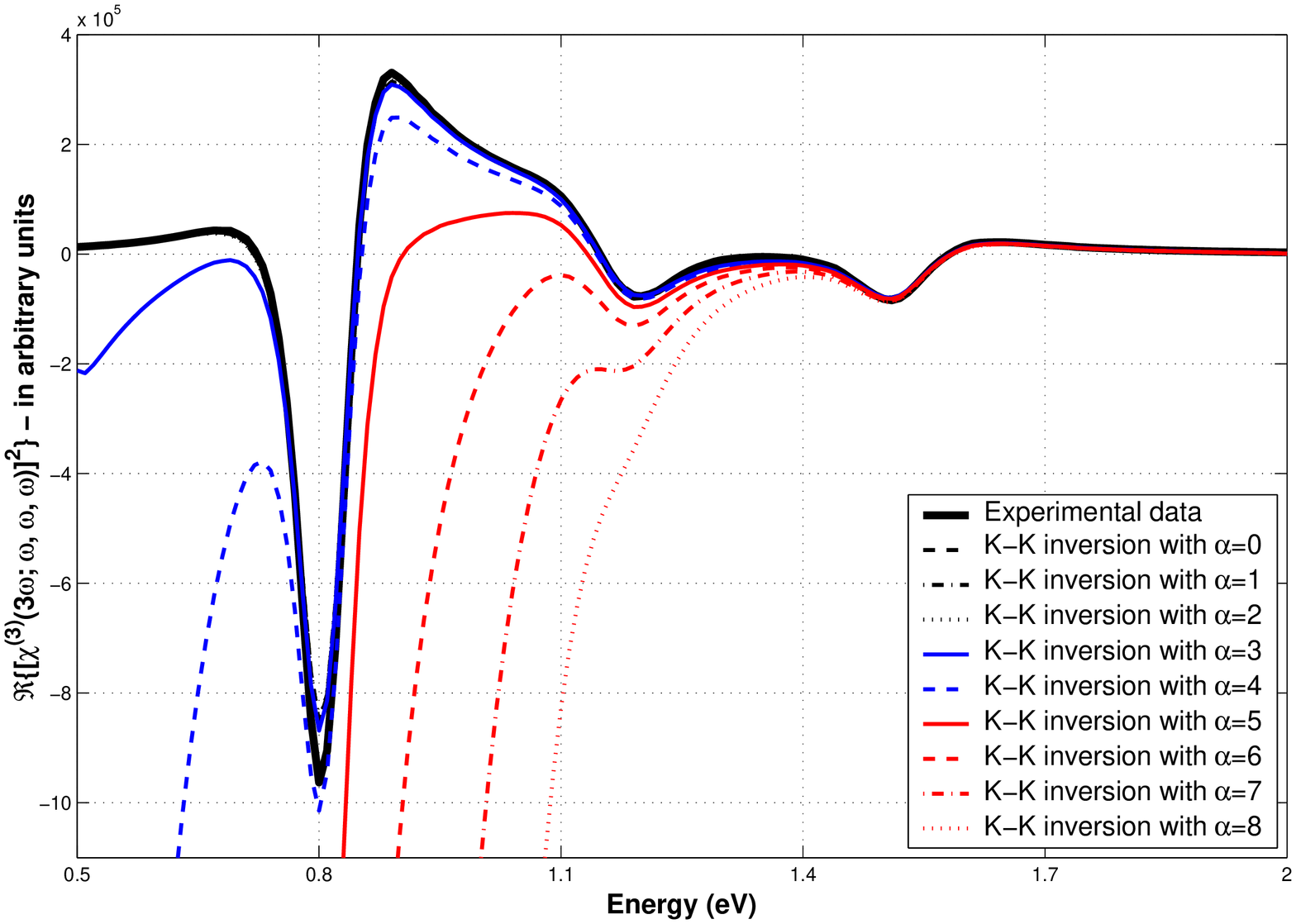}\\
  \caption{Lucarini, The Journal of Chemical Physics}\label{f7}
\end{figure}
\begin{figure}
  \includegraphics[angle=90, width=\textwidth]{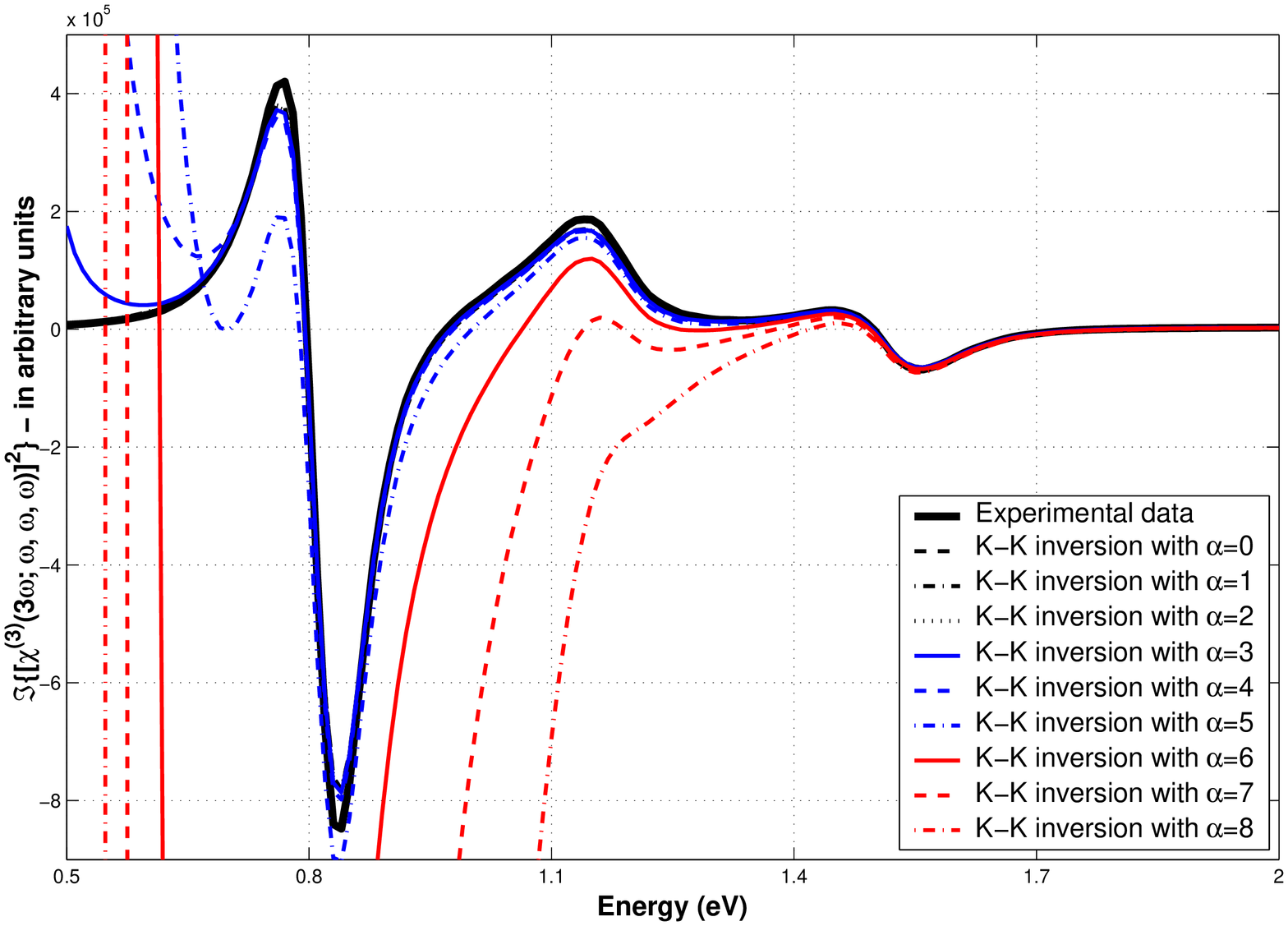}\\
  \caption{Lucarini, The Journal of Chemical Physics}\label{f8}

\end{figure}

In figures \ref{f7} and \ref{f8} we repeat the same analysis for
the second power of the susceptibility data taken on the
polythiophene: also in this case the agreement is very good if
$\alpha=0,1,2$, but the narrower frequency range does not allow
the data inversion if the very high moments are considered. If we
consider the real part -figure \ref{f7}- for $\alpha=3,4$ K-K data
inversion provides a good reproduction of the experimental data
for photon energies $\leq$ $0.7$ $eV$; for $\alpha\geq 5$ there is
no convergence in the lower half of the spectral range. For the
imaginary part -figure \ref{f8}- we can repeat the same
observations, except that for $\alpha=5$ there is still good
reproduction of the main features of the curve.
\par
We emphasize that if experimentally only one of the real or
imaginary part of the harmonic generation susceptibility has been
measured, there is no direct use of K-K relations relative to
higher powers of the susceptibility, since the multiplication
mixes the real and the imaginary parts. Therefore in this case the
K-K relations for $k>1$ can be used as tests of robustness of the
results obtained with the dispersion relations applied to the
conventional susceptibility.

\section{Verification of Sum Rules}\label{sect4}

Sum rules for optical functions which obey K-K relations can be
generally obtained by combining the dispersion relations with the
knowledge of their asymptotic behavior, which can be obtained with
a detailed analysis of the physics of the system at microscopic
level, by applying the superconvergence theorem
\cite{Nussenzveig,Altarelli1,Altarelli3} to the dispersion
relations.  \par In the case of arbitrary order harmonic
generation processes, we have that for large values of angular
frequency $\chi^{(n)}(n\omega)\approx\psi\omega^{-2n-2}$
\cite{ValerioHm,ValerioH}, where $\psi$ is a material-dependent
constant. Therefore from the K-K relations (\ref{k3}) and
(\ref{k4}) it is possible to derive the following sum rules
\cite{jarkko}:

\begin{equation}\label{k7}
\int_{0}^{\infty}\omega^{2\alpha}\Re\{[\chi^{(n)}(n\omega)]^{k}\}{\rm{d}}\omega=0,
 0\leq\alpha\leq k(n+1)-1
\end{equation}
\begin{equation}\label{k8}
\int_{0}^{\infty}\omega^{2\alpha+1}\Im\{[\chi^{(n)}(n\omega)]^{k}\}{\rm{d}}\omega=0,
0\leq\alpha\leq k(n+1)-2
\end{equation}
\begin{equation}\label{k9}
\int_{0}^{\infty}\omega^{2k(n+1)-1}\Im\{[\chi^{(n)}(n\omega)]^{k}\}{\rm{d}}\omega=-\frac{\pi}{2}\psi^{k},
\end{equation}

The verification of linear sum rules
\cite{Altarelli1,Altarelli2,Shiles et al. 1980,Altarelli3} from
experimental data is usually hard to obtain because of the
critical contributions given by the out-of-range asymptotic part
of the real or imaginary part of the susceptibility under
examination \cite{Fano,Bachelet}; however in the case of linear
optics information of the response of the material to very high
frequency radiation can be obtained with synchrotron radiation
\cite{Altarelli3}. In general a good accuracy in the verification
of sum rules is more difficult to achieve than for K-K relations,
therefore a positive outcome of this test  provide a very strong
argument to support the quality and the coherence of the
experimental data.
 \par In the case of harmonic nonlinear processes the technical limitations for achieving
information for a very wide frequency range are very severe, and
the verification of the sum rules is critical, especially for
those involving relatively large values of $\alpha$ which
determine a slower asymptotic decrease. Nevertheless, if we
consider increasing values of $k$, the integrands in the equations
(\ref{k7}), (\ref{k8}), and (\ref{k9}) have a much faster
asymptotic decrease, so that the missing high-frequency tails tend
to become negligible. Therefore we expect that for a given
$\alpha$ the convergence of the sum rules should be more accurate
for higher values of $k$, if we assume that the main features of
the spectrum are well reproduced by the experimental data, as
explained in the previous section.
\par
 We first focus on the vanishing sum rules (\ref{k7})-(\ref{k8}). In order to have a measure of how
precisely the vanishing sum rules are obeyed for the two
experimental data sets under examination, we introduce the
dimensionless quantities $Z_{\Re}$ and $Z_{\Im}$:

\begin{equation}\label{k5}
Z_{\Re}(\alpha,k)=|\frac{\int_{\omega_{min}}^{\omega_{max}}\omega^{2\alpha}\Re\{[\chi^{(3)}(3\omega)]^{k}\}{\rm{d}}\omega}
{\int_{\omega_{min}}^{\omega_{max}}\omega^{2\alpha}|\Re\{[\chi^{(3)}(3\omega)]^{k}\}|{\rm{d}}\omega}|,
\end{equation}
\begin{equation}\label{k6}
Z_{\Im}(\alpha,k)=|\frac{\int_{\omega_{min}}^{\omega_{max}}\omega^{2
\alpha+1}\Im\{[\chi^{(3)}(3\omega)]^{k}\}{\rm{d}}\omega}
{\int_{\omega_{min}}^{\omega_{max}}\omega^{2\alpha+1}
|\Im\{[\chi^{(3)}(3\omega)]^{k}\}|{\rm{d}}\omega}|.
\end{equation}

Low values of $Z_{\Re}(\alpha,k)$ and $Z_{\Im}(\alpha,k)$ imply
that the negative and positive contributions to the corresponding
sum rule cancel out quite precisely compared to their total
absolute value. The two data sets of the polymers have quite
different performances in the verification of these sum rules.
\par We present in figures 9 and 10 the results obtained
with the data taken of polysilane by considering $1 \leq k \leq 5$
for respectively the sum rules of the real and the imaginary part:
we can draw very similar conclusions for both cases. We see that
generally for a given $\alpha$, we have a better convergence when
a higher $k$ is considered, with a remarkable increase in the
accuracy of the sum rules for $k\geq3$. Consistently with the
argument that the speed of the asymptotic behavior is critical in
determining the accuracy of the sum rule, we also have generally a
decrease in the quality of the convergence to zero when, for a
given $k$, one considers higher moments, thus increasing the value
of $\alpha$. Particularly impressive is the increase of
performance in the convergence of the sum rules of
$\chi^{(3)}(3\omega)$ for both the real ($2\alpha=0,2,4,6$) and
the imaginary part ($2\alpha+1=1,3,5$) when we consider $k=4,5$
instead of $k=1$: the values of $Z_{\Re}$ and $Z_{\Im}$ decrease
of more than three orders of magnitude in all cases considered.
\par In figures 11 and 12 we present the corresponding results
for the experimental data taken of polythiophene. Most of the sum
rules computed with this data set show a very poor convergence to
zero, because the corresponding $Z_{\Im}$ and $Z_{\Re}$ are above
$10^{-1}$; nevertheless we can draw conclusions similar to the
previous case in terms of change of the accuracy of the
convergence for different values of $k$ and $\alpha$; consistently
with the relevance of the asymptotic behavior, the precision
increases for increasing $k$ and for decreasing $\alpha$. But in
this case for a given $\alpha$ the improvement in the convergence
of the sum rules obtained by considering a high value of $k$
instead of $k=1$ is generally small, being in most cases the
decrease of $Z_{\Im}$ and $Z_{\Re}$ below or around an order of
magnitude.
\par
We observe that the bias between the performances of the two data
sets in the verification of the vanishing sum rules is extremely
large, not comparable to the discrepancies found in the analysis
of the K-K relations. We can guess that the worse performance of
the data on polythiophene can be mainly attributed to their less
complete representation of the relevant nonlinear electronic
transitions of the material; as previously stated, the data
extension is critical in the verification of sum rules. This
result is consistent with the previously presented slightly worse
performance of this data set in the K-K inversion of the second
power of $\chi^{(3)}(3\omega)$, where the relevance of the
out-of-range data is also quite prominent.
\par Finally, from equation \ref{k9} it is
possible to obtain a simple relation between the non vanishing sum
rules for $[\chi^{(n)}(n\omega)]^{k}$ and the $k^{th}$ power of
the non vanishing sum rule for the conventional susceptibility
$\chi^{(n)}(n\omega)$:
\begin{equation}\label{k11}
-\frac{2}{\pi}\int_{0}^{\infty}\omega^{2k(n+1)-1}\Im\{[\chi^{(n)}(n\omega)]^{k}\}{\rm{d}}\omega=[-\frac{2}{\pi}\int_{0}^{\infty}\omega^{2n+1}\Im\{\chi^{(n)}(n\omega)\}{\rm{d}}\omega]^{k}
\end{equation}
For both of the two data sets considered in our work these
relations do not hold for $1 \leq k \leq 5$. The equation
\ref{k11} relates the slowest converging sum rules for each $k$,
therefore it is reasonable to explain the poor performances of the
experimental data in reproducing this theoretical results with the
finite range of data under examination.

\section{Conclusions}\label{sect5}

In this study we have performed the first thorough analysis of
generalized K-K relations and sum rules on experimental data of
third harmonic generation. We have tested the consistency between
the theory and experimental data in the worst case, which however
is usually most typical, namely data on limited spectral range
without extrapolations beyond the measured range, and without any
knowledge of anchor points \cite{Ahrenkiel,Palmer,ValJarKai}. We
have considered two data sets of comparable spectral range
referring to independent measurements of
$\Re\{\chi^{(3)}(3\omega)\}$ and $\Im\{\chi^{(3)}(3\omega)\}$
 performed on two polymers, the polysilane
\cite{Kishida}, whose data span a frequency range of \textit{0.4 -
2.5 eV}  and the polythiophene \cite{Torruellas1}, whose data span
a frequency range of \textit{0.5 - 2.0 eV}
\par
We have inverted the optical data of the susceptibility using
applying truncated K-K relations with a self-consistent procedure
and have verified for the first time that K-K relations hold
consistently also for the moments of the susceptibility
$\omega^{2\alpha}\chi^{(3)}(3\omega)$ with $0\leq \alpha \leq 3$,
as predicted by the general theory
\cite{ValerioHm,ValerioH,Rapapa}; the two data sets show very
similar performances. The precision of the data retrieved with K-K
relations is good in the upper part of the spectrum for all
moments, while for $\alpha=2,3$ there is a disagreement with the
experimental data in the very low portion of the spectrum, mainly
due to the absence of data descriptive of the asymptotic behavior
of the optical functions; nevertheless the main features of the
data are well reproduced by the dispersion relations. The
agreement between the results of the dispersion relations for the
various moments of the susceptibility under examination provide
fundamental information on the robustness of the retrieved data
and on the self-consistency of the experimental data.
\par
We have repeated the same analysis for
$\omega^{2\alpha}[\chi^{(3)}(3\omega)]^{2}$, in order to give the
first experimental confirmation of recent theoretical findings
that predict that K-K relations hold for
$\omega^{2\alpha}[\chi^{(n)}(n\omega)]^{k}$ with $0\leq \alpha
\leq k(n+1)-1$, so in our case dispersions relations connect the
real and imaginary part of
$\omega^{2\alpha}[\chi^{(3)}(3\omega)]^{2}$ if $0\leq \alpha \leq
7$. We have found that for the experimental third harmonic
susceptibility measured on polysilane K-K relations hold with a
high degree of precision for all the functions
$\omega^{2\alpha}[\chi^{(3)}(3\omega)]^{2}$ with $0 \leq
\alpha\leq 6$; the agreement between retrieved and experimental
data is excellent for $0 \leq \alpha\leq 2$. In the case of
polythiophene data, we can present similar conclusions, except
that the agreement exists only for $0 \leq \alpha\leq 5$. The
disagreement between the theory and the experimental fact can be
safely attributed to the truncation occurring in the high
frequency range, which mostly affects the convergence of the
dispersion relations of the very high moments. K-K relations for
higher powers of the susceptibility cannot be directly applied if
measures on only one part of the susceptibility are available, but
provide additional tests that inform on the completeness and
self-consistency of measured and retrieved data.
\par
We have investigated the validity of the sum rules that can be
derived by combining the knowledge of the asymptotic behavior of
$\chi^{(3)}(3\omega)$ with the previously described sets of K-K
relations for the moments of $\chi^{(3)}(3\omega)$ and of its
powers. The theory predicts that the integration in the
semi-infinite positive $\omega$-domain of each even moments of the
real part of the $k^{th}$ power of $\chi^{(n)}(n\omega)$ up to the
$(2k(n+1)-2)^{th}$ gives $0$, and the same holds for the odd
moments of the imaginary part of the $k^{th}$ power of the
susceptibility up to the $(2k(n+1)-3)^{th}$; the only nonvanishing
sum rule is the given by the integration of the $(2k(n+1)-1)^{th}$
moment of the imaginary part. In our case we have considered all
the powers of $\chi^{(3)}(3\omega)$ up to the fifth. \par In order
to evaluate the performance of the sum rules that the theory
predicts to be vanishing, we have introduced an dimensionless
measure of how precisely a sum rule computed on the experimental
data converges to zero. Generally, for both the sum rules of the
real and of the imaginary part, for a given $\alpha$, the
convergence improves if we consider a higher value of $k$.
Moreover, a decrease in the precision of the sum rules is realized
if for a given $k$, higher moments are considered. The main reason
for this behavior is in the concept that the faster the asymptotic
decrease of the integrand of the sum rule under examination, the
smaller the error due to the high-frequency cut-off. The two data
sets for polymers differ greatly in the precision achieved in the
verification of the sum rules, in many correspondent cases the
polysilane data provide results that are better by orders of
magnitude. The main reason for this discrepancy, much more
relevant but coherent with the results obtained in the  K-K study
of $[\chi^{(3)}(3\omega)]^{2}$, is the much stronger dependence of
the sum rules precision on the position of high frequency range
experimental cut-off relative to the saturation of the electronic
transitions of the material: it is likely that the data on
polythiophene are, apart from being narrower in absolute terms,
less comprehensive relatively to the electronic properties of the
material.
\par
On the contrary, the experimental data on both
polymers do not verify the non vanishing sum rules, which the
theory predicts to give information about the structure of the
material, because they do not obey the newly established
consistency relation (\ref{k11}) which should hold for each value
of the power $k$ considered for the susceptibility. The non
vanishing sum rules involve the highest moments for which the
integrations converge, therefore they are most strongly affected
by the frequency range finiteness; improvements on the data range
are then necessary to expect to obtain verification of these sum
rules.
\par
The constraints here analyzed are in principle universal, since
they essentially derive from the principle of causality in the
response of the matter to the external radiation, and so are
expected to hold for any material; they provide fundamental tests
of self-consistency  that any experimental or model generated data
have to obey; similar tests of coherence can be performed for
other nonlinear optical processes, e.g.
pump-and-probe\cite{Kaibook,Cataliotti 97,Scandolo,Valerio}.
Verification of K-K relations and sum rules constitute unavoidable
benchmarks for any investigation that addresses the nonlinear
response of matter to radiation on a wide spectral range.

\section*{Acknowledgments}
The authors would like to express their cordial thanks to Dr.
Hideo Kishida (Department of Advanced Materials Science,
University of Tokyo) and Dr. Takao Koda (Professor emeritus,
University of Tokyo) for providing the measured optical data on
polysilane.

\newpage
\section*{Figure captions}
\textbf{Fig. 1}: Efficacy of K-K relations in retrieving
$\Re\{\chi^{(3)}(3\omega)\}$ on polysilane
\\
\textbf{Fig. 2}: Efficacy of K-K relations in retrieving
$\Im\{\chi^{(3)}(3\omega)\}$ on polysilane
\\
\textbf{Fig. 3}: Efficacy of K-K relations in retrieving
$\Re\{\chi^{(3)}(3\omega)\}$ on polythiophene
\\
\textbf{Fig. 4}: Efficacy of K-K relations in retrieving
$\Im\{\chi^{(3)}(3\omega)\}$ on polythiophene
\\
\textbf{Fig. 5}: Efficacy of K-K relations in retrieving
$\Re[\{\chi^{(3)}(3\omega)\}]^{2}$ on polysilane
\\
\textbf{Fig. 6}: Efficacy of K-K relations in retrieving
$\Im[\{\chi^{(3)}(3\omega)\}]^{2}$ on polysilane
\\
\textbf{Fig. 7}: Efficacy of K-K relations in retrieving
$\Re[\{\chi^{(3)}(3\omega)\}]^{2}$ on polythiophene
\\
\textbf{Fig. 8}: Efficacy of K-K relations in retrieving
$\Im[\{\chi^{(3)}(3\omega)\}]^{2}$ on polythiophene
\\
\textbf{Fig. 9}: Convergence to $0$ of the vanishing sum rules
$\omega^{2\alpha}\Re[\{\chi^{(3)}(3\omega)\}]^{k}$ with $1 \leq k
\leq5$; data on polysilane
\\
\textbf{Fig. 10}: Convergence to $0$ of the vanishing sum rules
$\omega^{2\alpha+1}\Im[\{\chi^{(3)}(3\omega)\}]^{k}$ with $1 \leq
k \leq5$; data on polysilane
\\
\textbf{Fig. 11}: Convergence to $0$ of the vanishing sum rules
$\omega^{2\alpha}\Re[\{\chi^{(3)}(3\omega)\}]^{k}$ with $1 \leq k
\leq5$; data on polythiophene
\\
\textbf{Fig. 12}: Convergence to $0$ of the vanishing sum rules
$\omega^{2\alpha+1}\Im[\{\chi^{(3)}(3\omega)\}]^{k}$ with $1 \leq
k \leq5$; data on polythiophene

\newpage

\clearpage
\clearpage
\clearpage
\begin{figure}
  \includegraphics[angle=90, width=\textwidth]{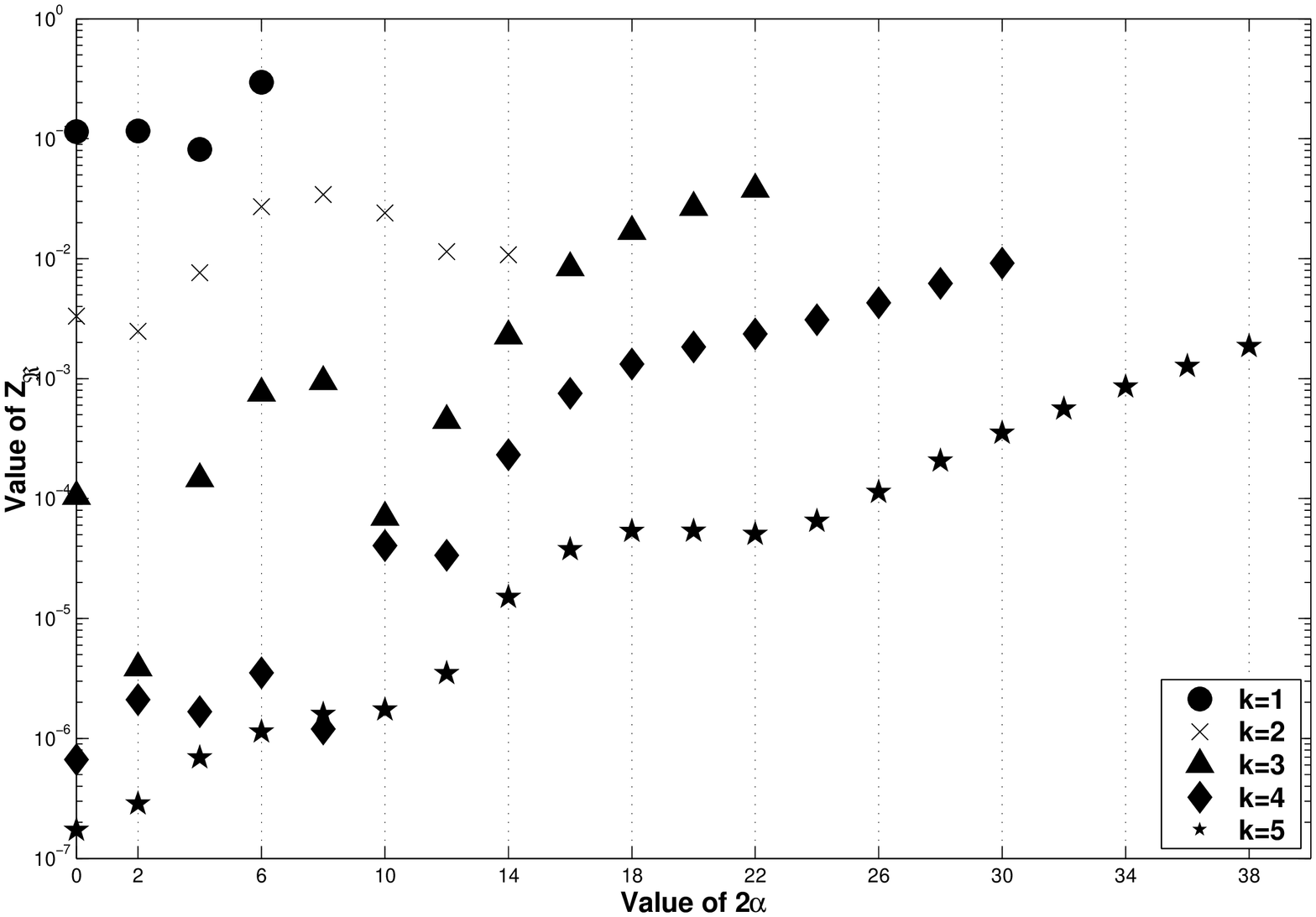}\\
  \caption{Lucarini, The Journal of Chemical Physics}\label{f9}
\end{figure}
\begin{figure}
  \includegraphics[angle=90, width=\textwidth]{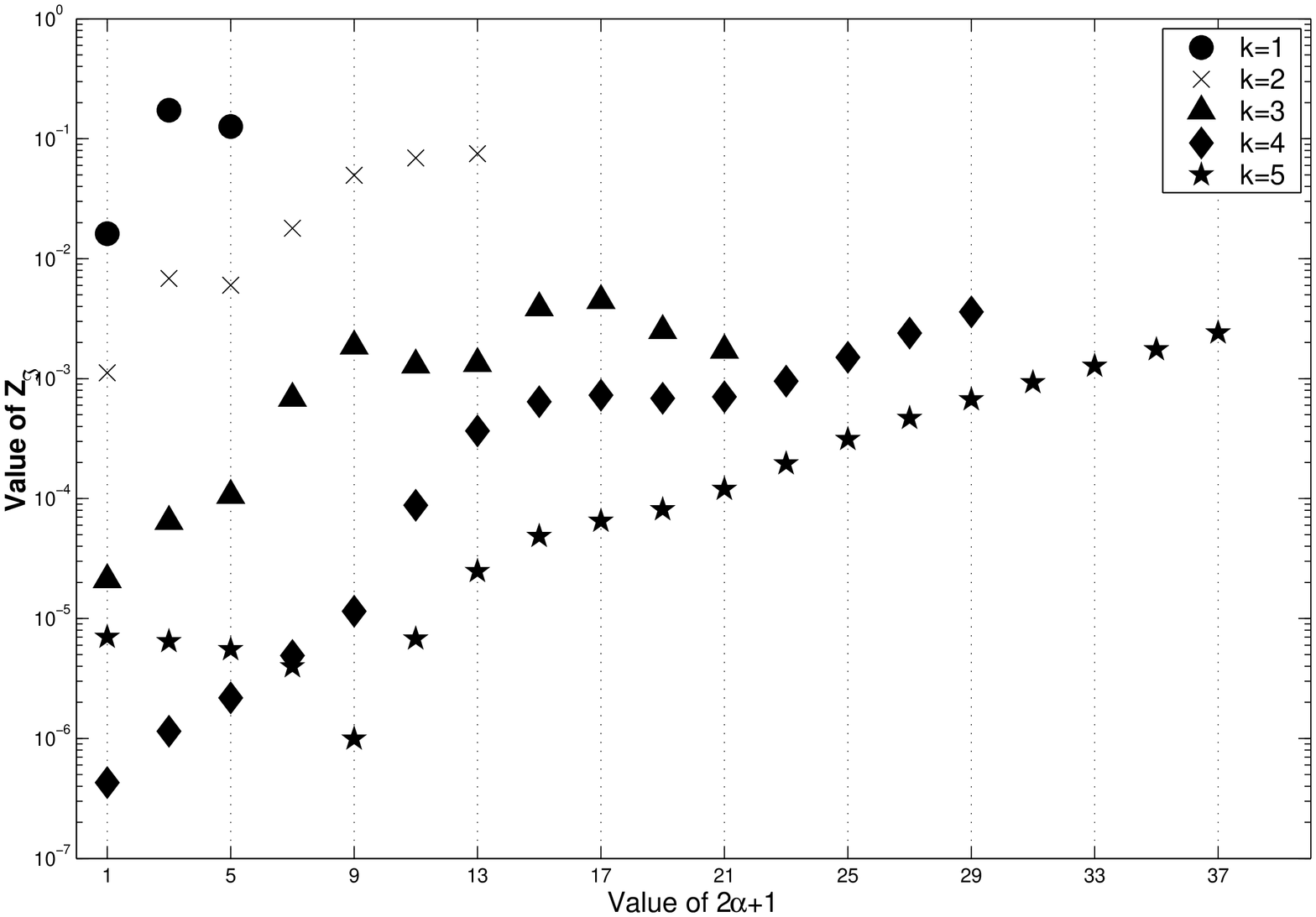}\\
  \caption{Lucarini, The Journal of Chemical Physics}\label{f10}
\end{figure}
\clearpage
\begin{figure}
  \includegraphics[angle=90, width=\textwidth]{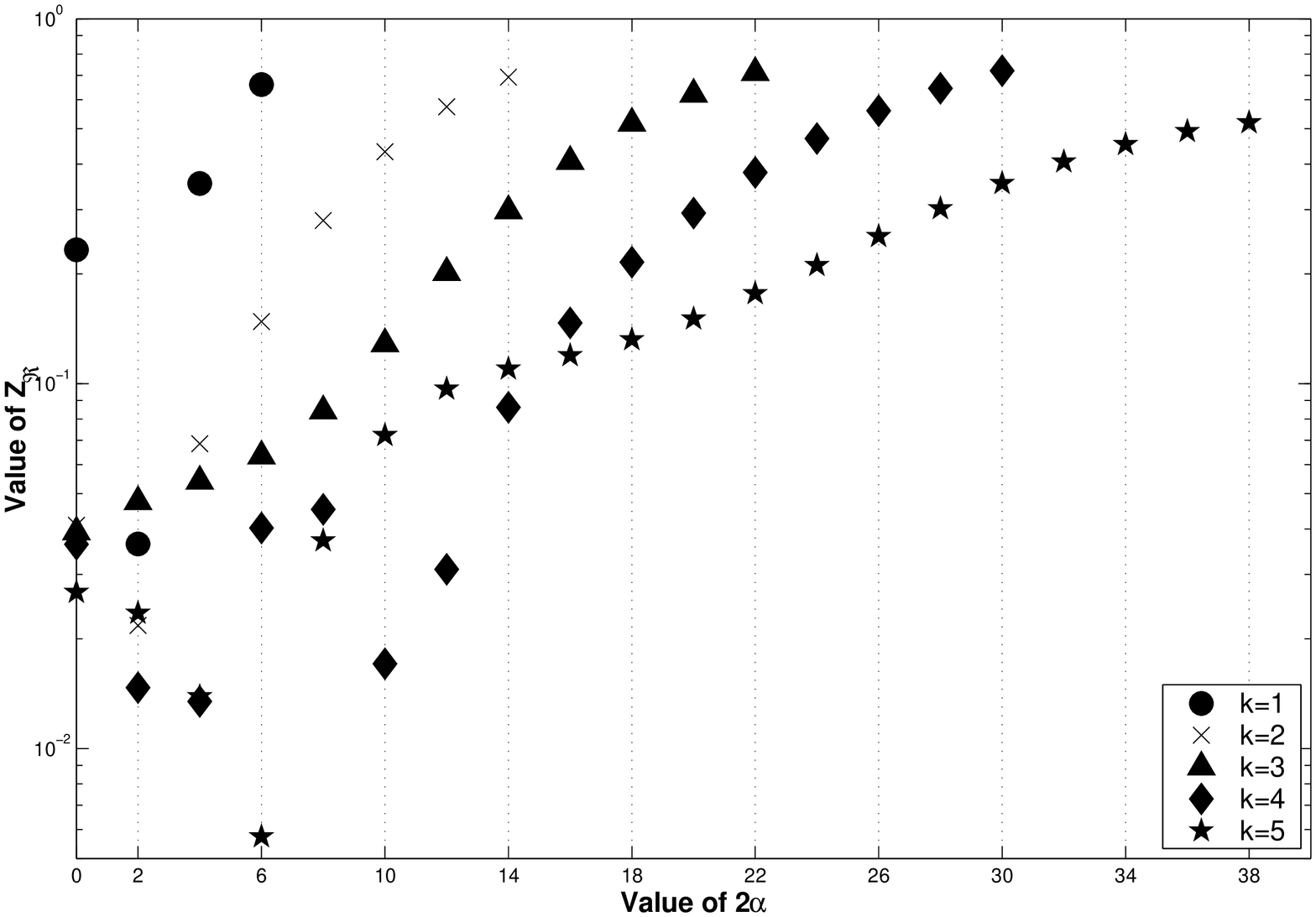}\\
  \caption{Lucarini, The Journal of Chemical Physics}\label{f11}
\end{figure}
\begin{figure}
  \includegraphics[angle=90, width=\textwidth]{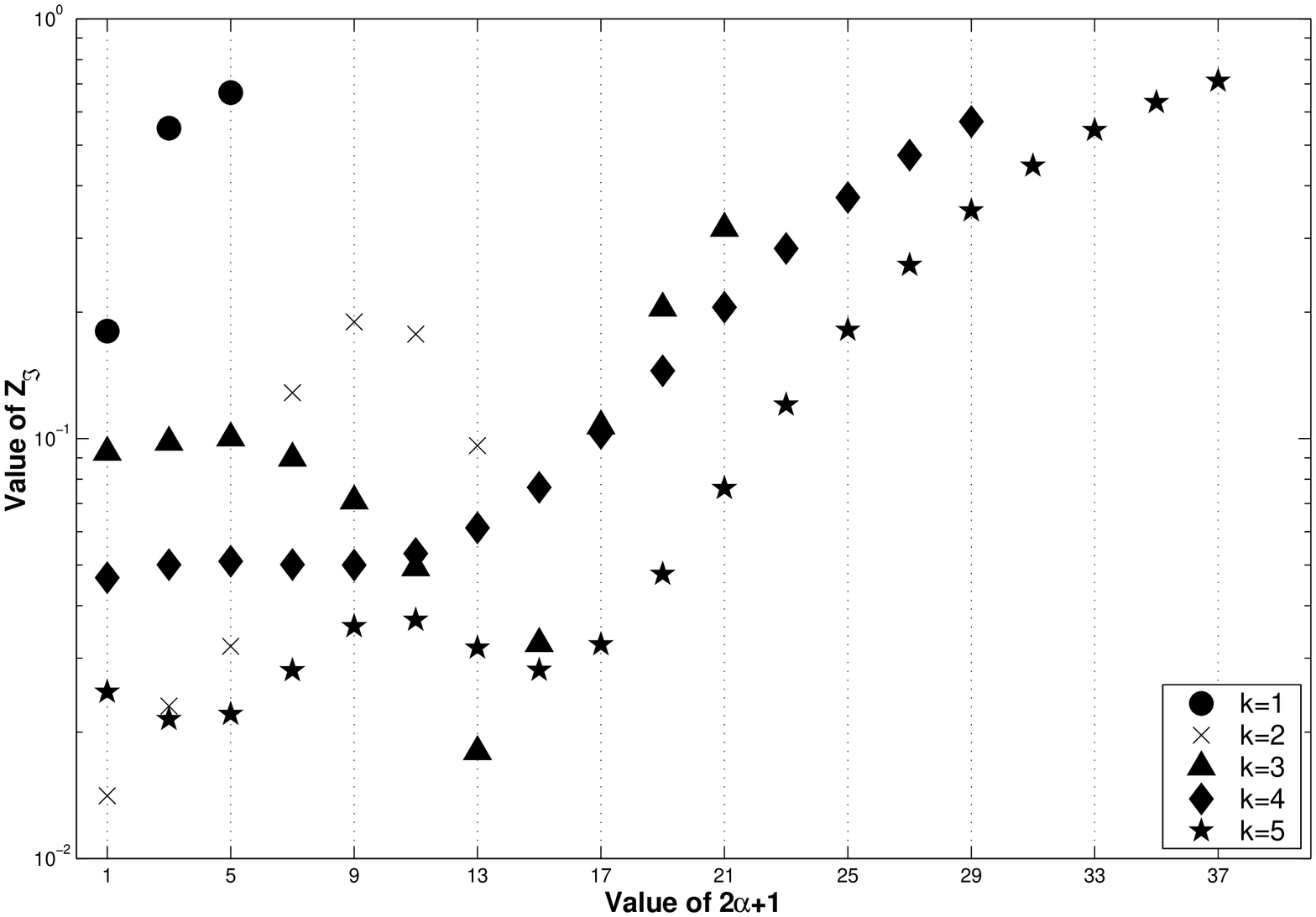}\\
  \caption{Lucarini, The Journal of Chemical Physics}\label{f12}
\end{figure}


\newpage

\begin{thebibliography}{[44]}

\bibitem{Landau}
L. D. Landau, E. M. Lifshitz, and P. Pitaevskii, {\em
Electrodynamics of Continuous Media} (Pergamon, Oxford, 1984).
\bibitem{Kaibook}
K.-E. Peiponen, E. M. Vartiainen, and T. Asakura, {\em Dispersion,
Complex Analysis and Optical Spectroscopy} (Springer, Heidelberg,
1999).
\bibitem{Altarelli1} M. Altarelli, D. L. Dexter, H. M. Nussenzveig, and D. Y. Smith, {\em Phys. Rev. B} {\bf 6},
4502 (1972)
\bibitem{Altarelli2} M. Altarelli and D. Y. Smith, {\em Phys. Rev. B} {\bf
9}, 1290 (1974)
\bibitem{Shiles et al. 1980} J. Shiles, T. Sasaki, M. Inokuti, and D. Y. Smith, \textit{Phys. Rev. B} \textbf{57}, 1612 (1980)
\bibitem{Grosso}G. Grosso and G. Pastori Parravicini, Solid State Physics (Academic Press, San Diego, Calif., 2000)
\bibitem{Altarelli3}F. Bassani and M. Altarelli, in \textit{Handbook of synchrotron radiation}, Vol. 1A p. 463, ed. E. E. Koch (North Holland, Amsterdam, 1983).
\bibitem{Kai3}K.-E. Peiponen and E. M. Vartiainen, {\em Phys. Rev. B} {\bf 44},
8301 (1991)
\bibitem{Aspnes} D. E. Aspnes, in \textit{Handbook of Optical Constants of
Solids}, Vol. I p. 89. (Academic, New York, 1998)
\bibitem{Milonni} P. W. Milonni, {\em J. Phys. B: At. Mol. Opt. Phys.} {\bf 35},
R31 (2002)
\bibitem{Nussenzveig} H. M. Nussenzveig, {\em Causality and Dispersion Relations} (Academic Press, New York,
1972)
\bibitem{Lee et al. 85}Y.H Lee, A. Chavez-Pirson, S.W. Koch, H.M.
Gibbs, S.H. Park, J. Morhange, A. Jeffrey, N. Peyghambarian, L.
Banyai, A.C. Gossard, and W. Wiegmann, \textit{Phys. Rev. Lett.}
\textbf{57}, 2446 (1985)
\bibitem{Cataliotti 97}F. S. Cataliotti, C. Fort, T. W. H\"{a}nsch, M. Inguscio, and M.
Prevedelli, \textit{Phys. Rev. A} \textbf{56}, 2221 (1997)
\bibitem{Gubler et al. 2000} U. Gubler, C. Bosshard, P. Günter, M.Y. Balakina, J. Cornil,
J.L. Brédas, R. Martin and F. Diederich, Proceedings of Conference
on Lasers and Electro-Optics (CLEO 2000), San Francisco,
California, USA, Technical Digest CMI1, 44 (2000).
\bibitem{Martinelli et al. 2000} M. Martinelli, L. Gomes, and R. J. Horowicz, \textit{Appl. Opt.} \textbf{39},
6193 (2000)
\bibitem{Price} P. J. Price, {\em Phys. Rev.} {\bf 130}, 1792  (1963)
\bibitem{Kogan} M. Kogan, {\em Sov. Phys. JETP} {\bf 16}, 217  (1963)
\bibitem{Caspers} W. J. Caspers, {\em Phys. Rev.} {\bf 133}, 1249 (1964)
\bibitem{Ghahramani et al. 1991} E. Ghahramani, D.J. Moss, J.E.
Sipe,\textit{ Phys. Rev. B} \textbf{43}, 9700 (1991)
\bibitem{Sheik et al. 92} M. Sheik-Bahae, D.J. Hagan and E.W. Van Stryland, \textit{Opt. and Quant.
Electr.}  \textbf{24}, 1 (1992)
\bibitem{Kador 95} L. Kador, \textit{Appl. Phys. Lett.} \textbf{66}, 2938 (1995)
\bibitem{Hughes et al. 1997} J. L. P. Hughes, Y. Wang, and J. E. Sipe, \textit{Phys. Rev. B} \textbf{55}, 13630
(1997)
\bibitem{Kai1} K.-E. Peiponen, {\em Phys. Rev. B} {\bf 35}, 4116  (1987)
\bibitem{Kai2} K.-E. Peiponen, {\em Phys. Rev. B} {\bf 37}, 6463  (1988)
\bibitem{Bassani1} F. Bassani and S. Scandolo, {\em Phys. Rev. B} {\bf 44},
8446  (1991)
\bibitem{Scandolo}S. Scandolo and F. Bassani, \textit{Phys. Rev. B} \textbf{45}, 13257 (1992).
\bibitem{Bassani2} S. Scandolo and F. Bassani, {\em Phys. Rev. B} {\bf 51}, 6925 (1995)
\bibitem{Kubo}R. Kubo, \textit{J. Phys. Soc. Japan} \textbf{12}, 570 (1957)
\bibitem{Peiponen02} K.-E. Peiponen and J. J. Saarinen, {\em Phys. Rev.
A} {\bf 65}, 063810  (2002)
\bibitem{Chernyak} V. Chernyak and S. Mukamel, {\em J. Chem. Phys.} {\bf 103},
7640  (1995)
\bibitem{ValerioHm} F. Bassani and V. Lucarini, {\em Il Nuovo Cimento D} {\bf 20}, 1117 (1998)
\bibitem{ValerioH} F. Bassani and V. Lucarini, {\em Eur. Phys. J. B} {\bf 17},
567  (2000)
\bibitem{Rapapa} N. P. Rapapa and S. Scandolo, {\em J. Phys.: Condens. Matter} {\bf 8}, 6997 (1996)
\bibitem{jarkko} J. J. Saarinen, {\em Eur. Phys. J. B} {\bf 30}, 551 (2002)
\bibitem{Fano} U. Fano and J. W. Cooper, \textit{Rev. Mod. Phys. }\textbf{40}, 441 (1968)
\bibitem{Bachelet} P. Alippi, P. La Rocca, and G.B. Bachelet \textit{Phys. Rev. B }\textbf{55}, 13835 (1997)
\bibitem{Ahrenkiel} R. K. Ahrenkiel, {\em J. Opt. Soc. Am.} {\bf 61}, 1651 (1971)
\bibitem{Palmer} K. F. Palmer, M.Z. Williams, and B.A. Budde, {\em Appl. Opt.} {\bf 37}, 2660 (1998)
\bibitem{ValJarKai} V. Lucarini, J. J. Saarinen, K.-E. Peiponen,
\textit{Opt. Commun.} (in press)
\bibitem{Torruellas1} W. E. Torruellas, D. Neher, R. Zanoni, G. I. Stegeman, F. Kajzar, and M. Leclerc, {\em Chem. Phys.
Lett.} {\bf 175}, 11 (1990)
\bibitem{Torruellas2} D. Guo, S. Mazumdar, G. I. Stegeman, M. Cha, D. Neher, S. Aramaki, W. Torruellas, and R. Zanoni, {\em Mater. Res. Soc. Symp. Proc.} {\bf 247}, 151 (1992)
\bibitem{Kishida} H. Kishida, T. Hasegawa, Y. Iwasa, T. Koda, and Y. Tokura, {\em Phys. Rev. Lett.} {\bf 70}, 3724 (1993)
\bibitem{Kishida2} H. Kishida, M. Ono, K. Miura, H. Okamoto, M. Izumi, T. Manako, M. Kawasaki, Y.Taguchi, Y. Tokura, T. Tohyama, K. Tsutsui, and S. Maekawa, \textit{Phys.
Rev. Lett.} \textbf{87}, 177401-1 (2001)
\bibitem{Valerio} F. Bassani and V. Lucarini, {\em Eur. Phys. J. B} {\bf 12},  323 (1999)






\end{thebibliography}
\end{document}